\documentclass[showpacs,prl,aps,superscriptaddress,twocolumn,floatfix]{revtex4}
\usepackage[english]{babel}
\usepackage{amsmath}
\usepackage{times}
\usepackage{amssymb,mathrsfs}
\usepackage{graphicx,color}
\definecolor{Green4}{RGB}{0,139,0}

\begin{document}

\title{Quantum Signature of Analog Hawking Radiation in Momentum
  Space}

\author{D. Boiron}\affiliation{Laboratoire Charles Fabry, Institut
d'Optique-CNRS-Univ. Paris-Sud, 2 avenue Augustin Fresnel, 91127
Palaiseau, France}

\author{A. Fabbri}\affiliation{Centro Studi e Ricerche E. Fermi,
  Piazza del Viminale 1, 00184 Roma, Italy} \affiliation{Dipartimento
  di Fisica dell'Universit\`a di Bologna, Via Irnerio 46, 40126
  Bologna, Italy} \affiliation{Departamento de Fisica Teorica and
  IFIC, Universidad de Valencia-CSIC, C. Dr. Moliner 50, 46100
  Burjassot, Spain} \affiliation{Laboratoire de Physique Th\'eorique,
  CNRS UMR 8627, B\^at. 210, Univ.  Paris-Sud, 91405 Orsay Cedex,
  France}

\author{P.-\'E. Larr\'e}\affiliation{INO-CNR BEC Center and
  Dipartimento di Fisica, Universit\`a di Trento, Via Sommarive 14,
  38123 Povo, Italy}

\author{N. Pavloff}\affiliation{Laboratoire de Physique Th\'eorique et
  Mod\`eles Statistiques, CNRS, Univ. Paris Sud, UMR 8626, 91405
  Orsay Cedex, France} 

\author{C. I. Westbrook}\affiliation{Laboratoire Charles
  Fabry de l'Institut d'Optique, CNRS, Univ. Paris-Sud, Campus
  Polytechnique RD128 91127, Palaiseau, France}

\author{P. Zi\'n}\affiliation{National Centre for Nuclear Research,
  ul. Ho\.{z}a 69, 00-681 Warsaw, Poland}

\begin{abstract}
  We consider a sonic analog of a black hole realized in the
  one-dimensional flow of a Bose-Einstein condensate. Our theoretical
  analysis demonstrates that one- and two-body momentum distributions
  accessible by present-day experimental techniques provide clear
  direct evidence (i) of the occurrence of a sonic horizon, (ii) of
  the associated acoustic Hawking radiation and (iii) of the quantum
  nature of the Hawking process. The signature of the quantum behavior
  persists even at temperatures larger than the chemical potential.
\end{abstract}


\maketitle

Forty years ago S. W. Hawking \cite{Hawking} discovered that black
holes are not completely ``black'' as General Relativity predicts, but
emit particles in the form of thermal radiation at the characteristic
temperature $T_{\rm H}=\kappa/2\pi c_{\ell}$, where $c_{\ell}$ is the
speed of light and $\kappa$ the horizon's surface gravity (we use
units such that $\hbar=k_{\scriptscriptstyle B}=1$). This subtle quantum
me\-cha\-nical effect can be understood as arising from a
pair-production process in the near-horizon region, in which one
member of the pair gets trapped inside the black hole leaving the
other ``free'' to propagate outside and reach infinity.  Unfortunately,
it seems impossible to observe Hawking radiation in the astrophysical
context because in ordinary situations of gravitational collapse $T_{\rm
  H}$ is much lower than the temperature of the microwave background
radiation \cite{remark}.  Different scenarios have been proposed that
would allow the formation of low mass black holes with higher values
of $T_{\rm H}$, but they remain speculative. Among these are the
suggestions that mini black holes might have been seeded by density
fluctuations in the early Universe \cite{Carr:1975qj} or could be
formed at particle accelerators due to the existence of large extra
dimensions \cite{minibhs}.

\begin{figure}[h]
\begin{center}
\includegraphics*[width=0.999\linewidth]{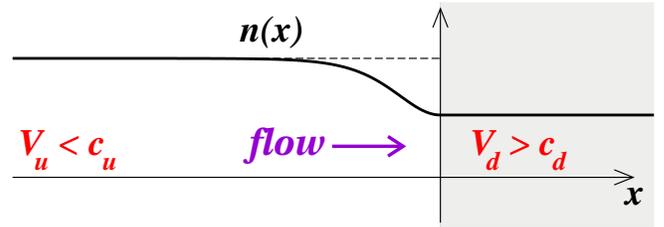}
\end{center}
\caption{(Color online) Schematic representation of an acoustic
  black-hole in a BEC. The density profile $n(x)$ is the black solid
  line. $V_u$ and $c_u$ ($V_d$ and $c_d$) are the asymptotic upstream
  (downstream) flow and sound velocities. The upstream (downstream)
  asymptotic flow is sub-sonic (super-sonic). The interior of the
  analog black hole is shaded in this figure and in
  Fig. \ref{fig.dispersion}.}
\label{BH.profile}
\end{figure}

In 1981 Unruh \cite{Unr81} used the mathematical equivalence between
the propagation of light in a gravitational black hole and that of
sound in a fluid undergoing a subsonic-supersonic transition
(henceforth denoted as an ``acoustic black hole'') to predict, using
Hawking's original analysis, that acoustic black holes will emit a
thermal flux of phonons (analog Hawking radiation) from their acoustic
horizon. Several physical systems have since been proposed to detect
the analog of Hawking radiation. Recent investigations attempted to realize
acoustic horizon in water tanks experiments \cite{Weinfurtner:2010nu},
{\it via} ultrashort pulses in optical fibers \cite{Phi08} or in a
transparent Kerr medium \cite{Belgiorno:2010wn}, by propagating
coherent light in nonlinear media \cite{Ela12}, in the flow of
micro-cavity polaritons \cite{Ngu15} and in atomic Bose-Einstein
condensates (BECs) \cite{Lah10}, see Fig.  \ref{BH.profile}.

Because of their low temperatures, BECs offer particularly
  favorable experimental conditions \cite{garayetal}, since one can
reach situations in which $T_{\rm H}$ is only one order of magnitude
lower than the background temperature in typical ultracold
atomic-vapor experiments. This is a significant improvement with
respect to the gravitational case, but it still seems too low to
attempt a direct detection of the emitted phonons. Fortunately,
acoustic black holes have another advantage compared to gravitational
ones: the interior of the analog black hole (region of
supersonic flow) is accessible to experiments. One can then test the
existence of the Hawking effect through the basic pair-production
process of Hawking quanta (in the exterior of the acoustic black
  hole) and of their partners (in the interior).  It was shown in
\cite{correlations} that this process features characteristic peaks in
the correlation function of the density fluctuations and that these
peaks exist in BECs. A signature of the Hawking effect,
amplified by a laser type instability \cite{BHlaser} in a black
hole-white hole setting, has been recently observed in
Ref. \cite{Stei14}. 

In the present Letter we consider new observables, namely the the one-
and two-body momentum distributions, and show that they yield a direct
signature of the Hawking effect and of its quantum nature. The
motivation for our approach comes from the recent experiment
\cite{Jas12} where momentum correlators were used to observe the
acoustic analogue of the dynamical Casimir effect \cite{Dyn_Casimir},
a pair creation process bearing strong analogies with the Hawking
effect, in which correlated particles are created in a homogeneous
system by a rapid temporal modulation of the system's Hamiltonian. The
momentum correlations are particularly interesting because, as shown
below, they offer a signature of the quantum nature of the Hawking
effect much less affected by the background temperature $T$ than the real
space correlation signal -- of intrinsically hydrodynamic
nature \cite{correlations} -- which has been recently studied in the
$T=0$ limit \cite{Stei15}.

\begin{figure}[h]
\begin{center}
\includegraphics*[width=0.999\linewidth]{dispersion}

\vspace*{-3mm}

\includegraphics*[width=0.75\linewidth]{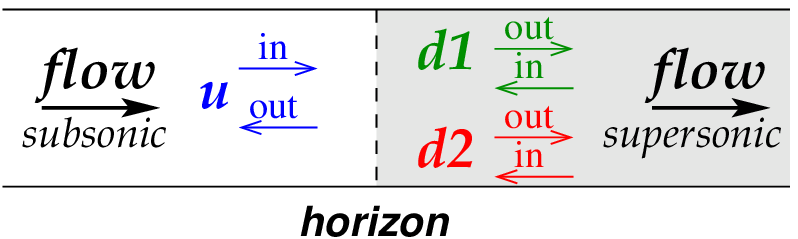}
\end{center}
\caption{(Color online) Two upper plots: upstream and downstream
  dispersion relations $\omega(k)$ [from Eq. \eqref{dispersion}]. In
  each plot the horizontal dashed line is fixed by the chosen value of
  $\omega$. The labeling of the modes is explained in the text; their
  direction of propagation is represented by an arrow. $u$ and $d1$
  modes ($d2$ modes) have positive (negative) norm. The $d2$ modes
    disappear for $\omega>\Omega$.  The lower plot schematically
  represents the black hole configuration in real space and the
  different modes bearing propagation of elementary excitations.}
\label{fig.dispersion}
\end{figure}

What we denote as an analog black hole is a stationary one-dimensional
(1D) flow in which the asymptotic upstream velocity is subsonic and
the asymptotic downstream velocity is supersonic. Such configurations, from
idealized to more realistic ones, have been proposed in
Refs. \cite{Leo03,correlations,Mac09,Rec09,Zap11,Lar12}. It has been
experimentally demonstrated in Ref. \cite{Lah10} and theoretically
shown in Ref. \cite{Kam12} how some could be reached by a
dynamical process. One of these configurations is schematically represented in
Fig. \ref{BH.profile}.  The sonic horizon is the place where the
velocity of the flow equals the speed of sound.

In such a structure, the dynamics of elementary excitations is encoded
in a $S$-matrix that describes how modes incoming from infinity
(upstream or downstream) are scattered by the horizon
\cite{Mac09,Rec09}. Far from the horizon the flow is uniform (with
constant velocity and density) and the distant incoming and outgoing
modes are thus plane waves. Their lab-frame dispersion relations are
of the Bogoliubov type (for the excitations propagating on top of a
uniform condensate, see, .e.g., \cite{Pit03}), Doppler shifted by the
background flow velocity:
\begin{equation}\label{dispersion}
(\omega-V_{(u/d)} k)^2 = 
c_{(u/d)}^2 k^2 \Big(1+ \tfrac{1}{4}\, k^2 \xi_{(u/d)}^2\Big) \; .
\end{equation}
In this expression $\omega$ is the frequency of the plane wave, $k$
its momentum relative to the background flow and $\xi_{(u/d)}=1/m
c_{(u/d)}$ is the (upstream or downstream) healing length. The
dispersion relations \eqref{dispersion} are represented in
Fig. \ref{fig.dispersion} where upstream and downstream modes are
denoted as ``$u$'' and ``$d$''. We follow the conventions of
Refs. \cite{Rec09,Lar12} and label the modes as ``in'' (such as, for
instance, $d1_{\rm in}$) or ``out'' (such as $u_{\rm out}$) depending
on whether their group velocity points toward the horizon (for the
``in'' modes) or away from the horizon (for the ``out'' modes), as
pictorially described in the lower part of
Fig. \ref{fig.dispersion}. In the upstream subsonic region the
dispersion relation is qualitatively similar to that of a
condensate at rest, with one ingoing and one outgoing $u$ channel;
new modes appear in the downstream supersonic region where we
have two ingoing and two outgoing modes, denoted as $d1$ and
$d2$. The new $d2$ modes have negative norm (see, e.g., \cite{Fet98}).
From this analysis one can identify the three relevant scattering
channels (each is initiated by one of the three ingoing modes)
and compute the coefficients of the $S$-matrix
\cite{Rec09,Lar12}. This, in turn, makes it possible to expand the
creation and annihilation operators $\hat\psi^\dagger(p)$
and $\hat\psi(p)$ on the scattering channel operators and to determine
the population operator $\hat{n}(p)=\hat\psi^\dagger(p)\hat\psi(p)$ of
the state with lab-frame momentum : $p = k+m V_{(u/d)}$,
where $k$ is the relative momentum of Eq. \eqref{dispersion}.

It is important to present the experimental detection scheme used for
measuring the momentum distribution, because this precisely defines
how the quantities described in the present Letter should be
theoretically evaluated.  The detection employed in Ref. \cite{Jas12}
that motivates our approach consists of opening the trap and letting
the elementary excitations be converted into particles expelled from
both ends of the condensate, according to a process known as ``phonon
evaporation'' \cite{Toz04}. As demonstrated in Ref. \cite{Jas12},
after an adiabatic opening of the trapping potential, a measure of the
velocity distribution of these particles gives access to the momentum
distribution $\langle \hat{n}(p)\rangle$ within the condensate and to
the correlator $g_2$ defined below in Eq. \eqref{g2}
\cite{other_remark}.

\begin{figure}[h]
\begin{picture}(8,4)
\put(0.35,0.35){\includegraphics*[width=8cm]{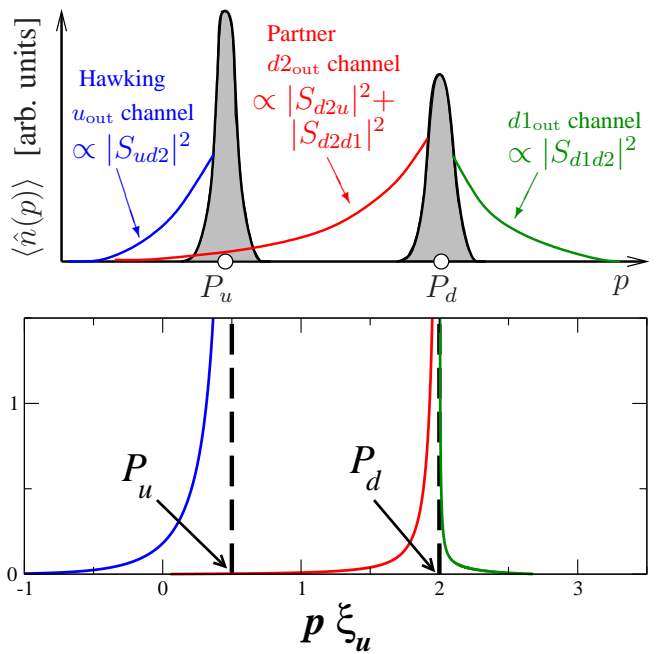}}
\put(-0.2,0.5){\rotatebox{90}{\large $\langle \hat{n}(p)\rangle$ 
\hspace{0.5 mm} [arb. units]}}
\put(2.3,0.0){\large $P_u$}
\put(5.3,0.0){\large $P_d$}
\put(7.8,0.1){\large $p$}
\put(0.7,2.8){\color{blue} Hawking}
\put(0.6,2.35){\color{blue} $u_{\rm out}$ channel}
\put(0.6,1.85){\color{blue} \large $\propto |S_{ud2}|^2$}
\put(1.2,1.7){\color{blue} \vector(1,-3){0.3}}
\put(3.2,3.4){\color{red} Partner}
\put(3.2,3.0){\color{red} $d2_{\rm out}$ channel}
\put(3.0,2.5){\color{red}\large $\propto |S_{d2u}|^2 +$}
\put(3.5,2.05){\color{red}\large $|S_{d2d1}|^2$}
\put(4.,1.85){\color{red}\vector(1,-2){0.3}}
\put(6.4,2.25){\color{Green4}$d1_{\rm out}$ channel}
\put(6.4,1.8){\color{Green4}\large $\propto |S_{d1d2}|^2$}
\put(6.8,1.65){\color{Green4} \vector(-1,-2){0.3}}
\end{picture}

\vspace{2mm}

\includegraphics*[width=0.99\linewidth]{onebody_theo}
\caption{(Color online) Momentum distribution $\langle \hat{n}(p)\rangle$
  within the condensate. The figure is drawn in the $T=0$ limit. The
  top panel is a schematic representation of a typical experimental
  result. The lower panel displays the analytic result in the
  waterfall configuration with $V_u/c_u=0.5$, $V_d/c_d=4=V_d/V_u$. The
  thick vertical dashed lines correspond to the upstream and
  downstream condensates located at $p=P_u$ and $p=P_d$ (where here
  $P_u\xi_u=0.5$ and $P_d \xi_u=2$). }\label{fig.onebody}
\end{figure}

Figure \ref{fig.onebody} displays the one-body momentum distribution
$\langle \hat{n}(p) \rangle$ corresponding to the above-defined
procedure in the $T=0$ limit. The top part of this figure sketches the
expected typical experimental result. The shaded peaks are the
upstream or downstream momentum distribution of the condensate,
centered around $P_u$ and $P_d$ ($P_{(u/d)}=m V_{(u/d)}$). The lower
part of the figure displays our theoretical results obtained within
the so-called ``waterfall configuration'' where the sonic horizon is
induced by an external potential step \cite{Lar12}. Very similar
results are obtained for another realistic configuration (denoted as
``$\delta$-peak configuration'' in Ref. \cite{Lar12}) where the
horizon is induced by a sharply localized potential.  In our
theoretical description, the system behaves as a perfect 1D BEC (see
the discussion at the end of the Letter). In this case the components
of the momentum distribution corresponding to the condensate (the
dashed lines at $p=P_u$ and $p=P_d$ in the lower part of
Fig. \ref{fig.onebody}) are sharp $\delta$ distributions; they are not
broadened by phase fluctuations and finite experimental resolution as
in the top panel.

It is noteworthy that the presence or absence of a horizon can
be inferred from the structure of the one-body momentum distribution
$\langle \hat{n}(p)\rangle$. As illustrated in Fig. \ref{fig.onebody},
when a horizon is present, one has two peaks with $P_d > P_u$. On the
other hand, without horizon, one always has $P_d\le P_u$, the equality
being realized in the $\delta$-peak configuration \cite{Usdetails}.

The side distributions around the peaks in Fig. \ref{fig.onebody} are
signatures of the quantum fluctuations and are proportional to the
elements of the $S$-matrix ($S_{ud2}$ is, for instance, the complex and
$\omega$-dependent scattering amplitude describing the scattering from
the ingoing downstream channel $d2_{\rm in}$ towards the outgoing
upstream channel $u_{\rm out}$). In particular, the left shoulder of
the peak around $P_{u}$ in Fig.~\ref{fig.onebody} corresponds to the
Hawking quanta escaping from the horizon along the $u_{\mathrm{out}}$
channel, and the left shoulder of the peak around $P_{d}$ to their
partners ($d2_{\mathrm{out}}$ channel) \cite{NotePartners}. At $T=0$, the
existence of these shoulders stems directly from the Hawking effect; they
disappear in the absence of the horizon.

We now consider the normally ordered momentum correlation function (two
body signal)
\begin{equation}\label{g2}
g_2(p,q)= \frac{\langle : \! \hat n(p)\, \hat n(q) \! : \rangle
}{\langle \hat n(p) \rangle \, 
\langle \hat n(q) \rangle} \; .
\end{equation}
\begin{figure}
\includegraphics*[width=0.99\linewidth]{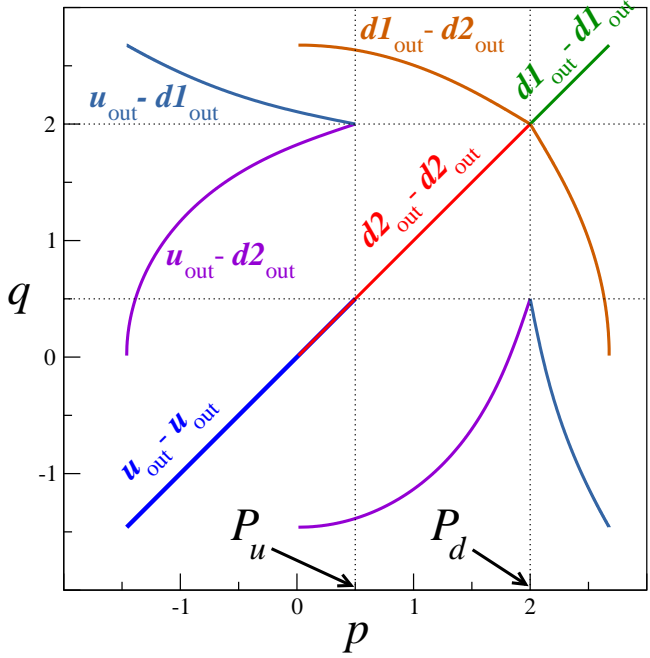}
\caption{(Color online) Momentum correlation pattern of $g_2(p,q)$ in
  the presence of a black hole horizon at $T=0$. This plot is drawn
  for the same configuration and the same parameters as the lower panel
  of Fig. \ref{fig.onebody}. The dotted lines are the momenta of the
  upstream and downstream condensates ($P_u$ and $P_d$). The momenta
  are expressed in units of $\xi_u^{-1}$.  Except for the colored
  correlation lines, $g_2(p,q)$ is uniformly equal to 1. The colors
  are used for a nonambiguous identification of the correlation
  lines. As is obvious from the definition \eqref{g2}, the figure is
  symmetric with respect to the diagonal.}
\label{zeroT}
\end{figure} 
Instead of presenting here the detailed analytical evaluation of $g_2$
corresponding to the different possible steps of the experimental
procedure that our theoretical approach is able to describe (see
Ref. \cite{Usdetails}), we rather graphically display $g_2$ at $T=0$
in Fig. \ref{zeroT}. This plot exhibits the genuine Hawking
correlations: in the absence of a horizon, the $T=0$ normal-ordered
$g_2$ would be uniformly equal to 1.  In our stationary setting,
spontaneous particle creation \`a la Hawking is triggered by the
existence of the negative norm $d2_{\rm in}$ mode. The process is
possible within an energy conserving framework because the d2 modes
carry a negative energy \cite{Bal10,Rob12}.  Hence the observation of
the new correlation lines $u_{\rm out}-d2_{\rm out}$, $d1_{\rm
  out}-d2_{\rm out}$ and $u_{\rm out}-d1_{\rm out}$ is a direct
evidence of the existence of the negative norm (negative energy) $d2$
modes and of a region of supersonic flow.  As discussed below, we work
within a perfect condensate approximation where the momenta are
exactly $\delta$-correlated along these curves.  Compared to the one
body signal displayed in Fig. \ref{fig.onebody}, the measure of the
momentum correlation function has the advantage of yielding a signal
located around easily identifiable curves.  These curves
-- and therefore the Hawking process -- terminate at momenta for which
the $d2$ modes disappear : this corresponds to the regime where
$\omega > \Omega$ in Fig. \ref{fig.dispersion}.

\begin{figure}
\includegraphics*[width=0.99\linewidth]{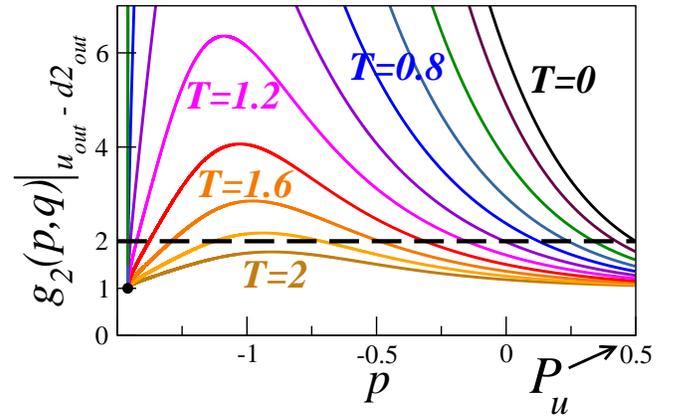}
\caption{(Color online) The value of $g_2(p,q)$ along the $u_{\rm
    out}-d2_{\rm out}$ correlation line (violet line in
  Fig. \ref{zeroT}) is plotted as a function of $p$ (expressed in
  units of $\xi_u^{-1}$) for different temperatures. The
  Cauchy-Schwarz inequality is violated when $g_2(p,q)$ is larger
  than 2. The temperatures are expressed in units of the upstream
  chemical potential $m c_u^2$ ($0\le T\le 2$). For the set of
  parameters chosen here and in the previous figures, the Hawking
  temperature is $T_{\rm H}=0.134$.}
\label{CS_T}
\end{figure}

Let us now turn to the quantitative study of the nature of
  correlations along the lines identified in Fig. 4. The
  occurrence of entanglement and the quantum nature of the Hawking
process can be tested through the
violation of the Cauchy-Schwarz inequality, as recently studied in a
similar context in Refs.  \cite{CSviolation}. More specifically, the
Cauchy-Schwarz inequality is violated along the characteristic Hawking
quanta$-$partner correlation lines $u_{\rm out} - d2_{\rm out}$
of Fig. \ref{zeroT} if (see, e.g., \cite{WallsMilburn})
\begin{equation}\label{CS}
g_2(p,q)\Big|_{u_{\rm out}-d2_{\rm out}}
>
\, \sqrt{g_2(p,p)\Big|_{u_{\rm out}}\!\!\times
\, g_2(q,q)\Big|_{d2_{\rm out}} } \; .
\end{equation}
In Fig. \ref{CS_T} this corresponds to the region located above the
dashed horizontal black line. In the Bogoliubov approach used in
  the present Letter, Wick's theorem yields
  $g_{2}(p,p)|_{u_{\mathrm{out}}} = g_{2}(q,q)|_{d2_{\mathrm{out}}}=2$
  for all temperatures; as a result, the right-hand side of inequality
  \eqref{CS} is equal to 2. The computations are
done in a setting where the system is in an initial thermal state at
temperature $T\ne 0$, and where the population of quasiparticles is
adiabatically converted into real particles upon opening the trap. As
expected the region of violation of the Cauchy-Schwarz inequality
decreases when $T$ increases \cite{remark_hydro}, but even at
relatively high temperatures ($T> 1.5\, m c_u^2$ and $>10\, T_{\rm H}$)
the momentum correlation signal remains a clear signature for
revealing the quantum nature of the Hawking signal
  \cite{rem_entang}.

Finally, it is important to stress that the results presented in this
work are obtained within Bogoliubov approximation assuming
  perfect condensation of the 1D Bose system. This approximation is
valid in an intermediate density regime -- denoted as
``1D mean field'' in Ref. \cite{Men02} -- where the system is
accurately described by an order parameter obeying an effective
1D Gross-Pitaevskii equation. At low density, phase fluctuations
destroy the long range order and the possibility of a true
Bose-Einstein condensate, and blur the sharp correlations
of Fig. \ref{zeroT} \cite{QCond_correl}.  At large density,
phase fluctuations can be neglected, but one cannot omit the effect of
transverse confinement which induces a modification of the dispersion
relation and creates new transverse dispersion modes
\cite{transverse}, resulting in the appearance of new correlation lines in
  Fig. \ref{zeroT}.  One should, however, keep in mind that for a
typical system (say, $^{87}$Rb, $^{23}$Na or $^4$He atoms in a guide
with a transverse confinement of angular frequency
$\omega_\perp=2\pi\times 500$ Hz) the 1D mean field approximation used
in the present Letter is quite relevant because it holds for a range of
linear densities varying over 4 orders of magnitude \cite{remark4}.

\begin{acknowledgments}
  This work was supported by the French ANR under Grant
  No. ANR-11-IDEX-0003-02 (Inter-Labex grant QEAGE), by the Triangle
  de la Physique, and by the Institut Francilien pour la Recherche en
  Atomes Froids. P. Z. was supported by the National Science Centre
  Grant No. DEC-2011/03/D/ST2/00200.
\end{acknowledgments}


\begin{thebibliography}{99}
\bibitem{remark} From the surface gravity formula, one obtains $T_{\rm H}\simeq
  10^{-7} M^{-1}$ K, where $M$ is the mass of the black hole expressed in
  units of the mass of the Sun.
\bibitem{other_remark} One can also imagine performing measurements of
  the momentum of the excitations by using two consecutive Bragg
  pulses as proposed in Ref. \cite{Bru00} and performed in
  Ref. \cite{Vog02}. However, it seems challenging to precisely
  measure correlations with this procedure.
\bibitem{NotePartners} After removing the external
    potentials, the acoustic horizon disappears, the collective
    excitations correspondingly transform into real particles, and the
    Hawking partners (within the $d2_{\mathrm{out}}$ channel)
    propagate with a lab-frame momentum $p=P_{d}-k$ smaller than
    $P_{d}$, thus contributing to $\langle\hat{n}(p)\rangle$ with a
    shoulder at the left of $P_{d}$ in Fig.~\ref{fig.onebody}.
\bibitem{remark_hydro} In particular, even for small temperatures, the
  Cauchy-Schwarz inequality is not violated in the hydrodynamic
  $p\,\xi_u \sim 0.5$ regime or in the region $p\,\xi_u \simeq -
  1.46$ where Hawking radiation shuts off and is easily superseded by
  thermal effects.
\bibitem{rem_entang} Entanglement is also present along the $d1_{\rm
    out}-d2_{\rm out}$ correlation lines (but is less easily detected)
  and absent along the $u_{\rm out}-d1_{\rm out}$ lines.
\bibitem{remark4} The 1D mean field regime holds, at $T=0$, for linear
  densities $n$ satisfying $(a/a_\perp)^2 \ll n \, a \ll 1$ where
  $a_\perp=(\hbar/m\omega_\perp)^{1/2}$ and $a$ is the 3D s-wave
  scattering length. In the examples chosen in the text the ratio
  $a/a_\perp$ is roughly of order $10^{-2}$ and $n$ should thus satisfy
$10^{-4}\ll n\,a \ll 1$.
\end{thebibliography}
\end{document}